# Understanding the merging behavior patterns and evolutionary mechanism at freeway on-ramps

Yue Zhang
Key Laboratory of Road and Traffic Engineering of Ministry of Education
Tongji University, Shanghai 201804, China
EMail: zhangyue18@tongji.edu.com

Yajie Zou*
Associate Professor
Key Laboratory of Road and Traffic Engineering of Ministry of Education
Tongji University, Shanghai 201804, China
EMail: yajiezou@hotmail.com

Lingtao Wu
Assistant Research Scientist
Texas A&M Transportation Institute
Texas A&M University System, 3135 TAMU, College Station, Texas 77843-3135
Email: wulingtao@gmail.com

Wanbing Han
Associate Professor
Key Laboratory of Road and Traffic Engineering of Ministry of Education
Tongji University, Shanghai 201804, China
EMail: 2131365@tongji.edu.cn

Word Count: 2465 words + 5 table (250 words per table) = 2,715 words

*Submitted [Submission Date]*

**ABSTRACT**
Understanding the merging behavior patterns at freeway on-ramps is important for assistanting the decisions of autonomous driving. This study develops a primitive-based framework to identify the driving patterns during merging processes and reveal the evolutionary mechanism at freeway on-ramps in congested traffic flow. The Nonhomogeneous Hidden Markov Model is introduced to decompose the merging processes into primitives containing semantic information. Then, the time-series K-means clustering is utilized to gather these primitives with variable-length time series into interpretable merging behavior patterns. Different from traditional state segmentation methods (e.g. Hidden Markov Model), the model proposed in this study considers the dependence of transition probability on exogenous variables, thereby revealing the influence of covariates on the evolution of driving patterns. This approach is evaluated in the merging area at a freeway on-ramp using the INTERACTION dataset. Results demonstrate that the approach provides an insight about the complicated merging processes. The findings about interpretable merging behavior patterns as well as the evolutionary mechanism can be used to design and improve the merging decision-making for autonomous vehicles.

**INTRODUCTION**

The freeway on-ramp is an important roadway segment, which often causes recurrent bottleneck. Due to the unique geometric characteristics of the on-ramps, the driver is often forced to perform a merging behavior to enter the mainline. In the process of merging, the driver needs to not only interact closely with the surrounding vehicles, but also observe other factors such as the driving environment. Merging at freeway on-ramps is an indispensable but challenging task, which has considerable consequences for traffic flow efficiency and driver safety *(1)*. To complete such complex tasks, human drivers need to constantly play games with the surrounding environment during the merging process, and react quickly based on their perception and experience. However, the existing autonomous vehicle decision-making only relies on models and explicit rules *(2)*, but lacks human insight and analysis. This is one of the core reasons why the autonomous vehicles cannot adequately handel the complex scenarios encountered in the real world. Therefore, to expand the prior knowledge of autonomous driving decision-making system, it is necessary to thoroughly investigate the merging behavior patterns and its evolution mechanism in the real world.

Numerous studies have been conducted to analyze merging behaviors. One common method to describe merging behavior is gap acceptance theory *(3-5)*. This technique assumes that vehicles will execute merging maneuvers if the gap between merging vehicles and the lead/lag vehicles in the target lane is acceptable. As the simplicity of the model, it is widely used in simulation software packages such as Vissim and Aimsun. It should be noted that the drivers may still merge even if the current gap value is smaller than critical gap in reality. Other studies classified the merging behaviors into several categories based on different criteria. According to the gap acceptance theory, the single merging process is divided into two stages: selecting the acceptable gap and implementing the lane change *(6)*. Considering the interaction between the merging vehicle and the surrounding vehicles, Sun et al. *(7)* categorized merging behavior into three classes: normal, forced and cooperative. Chu et al. *(8)* classified vehicle merging behaviors as direct merging, yield merging, and chase merging. The researchers further used discrete choice models to explore the best model. The study assumes that vehicles make merging decisions at a fixed point, but in fact the decision points are not exactly the same for each merging event. Especially in a congested traffic flow, the driver may need to continuously make the decision whether or not to change lanes. However, the above studies ignored the dynamic decision-making process in the merging behavior, and analyzed the merging behavior at a coarse-grained level. In order to conduct a more refined analysis, a number of researchersconsidered modeling the dynamic merger behavior. Wan et al. *(9)* proposed a sequential choice model to dynamically simulate the merging tactics under the changing traffic conditions. Meng and Weng *(10)* developed an improved cellular automata model for simulating dynamic merging behaviors and analyzed the effect of variables on the merging process. A time-varying mixed logit regression model are developed to describe merging behavior *(11)*. The game theory was applied to make drivers' merge decisions by establishing expected utility models *(12; 13)*. Besides, there are some data-driven methods developed to capture time-varying merging behaviors and make real-time decisions, such as classification and regression tree *(14)*, dynamic Bayesian network *(15)*, and fuzzy logic models *(16)*, etc. These studies analyzed the merging behavior of vehicles at every moment (fine-grained), which is computationally intensive and difficult to explain the evolution law of driving behavior during the merging process.

One way to solve the above problems is to decompose the whole merge process into several driving primitives (medium-grained). Because the driving behavior may not change significantly in a short period of time, some reseachers segment the entire driving scenarios into discrete fragments. This method facilitates to gain an insight into what happens inside merging process *(17)* so as to provide prior knowledge for autonomous vehicle decision-making. And discretization can save the cost of calculation and promote the decision-making process of autonomous vehicles. Higgs and Abbas *(17)* developed a two-step algorithm for the segmentation and clustering of car-following behaviors, in which a segmentation optimization equation was used for segmentation and k-means was utilized for clustering. After that, the Hidden Markov Model (HMM) is developed as a common primitive segmentation method, which is suitable to mine dynamic internal state in sequences. In the beginning,

*Zhang et al.*

HMM is widely used in speech recognition such as lexical decoding, syntactic analysis, and semantic analysis *(18)*. Wang et al. *(19)* introduced the hidden semi-Markov model to extract the primitive driving patterns from car-following scenarios. Zhang and Wang *(20)* integrated a hierarchical Dirichlet process with a hidden Markov model to decompose the intersection driving behavior data into small interpretable components and the K-means method to gather these driving primitives to driving patterns. analyzed the evolution of lane-changing interactive pattern and the formation mechanism of risk based on the Hidden Markov Model with the Gaussian mixture model. Although these methods are able to explain the internal driving patterns well, they do not consider the influence of exogenous factors on the driving patterns. On freeway on-ramps, the merging behavior of the vehicle itself is affected by various factors such as gap *(22)*, relative speed, relative distance, distance to the start of the ramp *(4)*, whether a lead vehicle exists in the merge lane, the type of surrounding vehicles *(11)*, and traffic density *(23)*, etc. Ignoring the influence of exogenous factors on internal states may not only lead to inaccurate driving primitive segmentation, but also hinder understanding on the evolution mechanism of driving patterns. Finally, it will also lead to the failure of autonomous vehicles to make appropriate decisions.

Based on the previous studies, the dynamic merging behavior patterns in the congested traffic flow still need to be explored an interpreted semantically. Also, it is necessary to propose an approach to explain how behavior patterns evolve and how exogenous variables drive the evolution of behavior patterns in the merging process.

This study develops a primitive based framework to understand the underlying merging behavior patterns for potential decision-making applications.

The main contributions of this study are as follows:

1) Proposing a medium-granularity analysis framework to identify and interpret the merging behavior patterns on the highway on-ramps in congested traffic flow.

2) Introducing the Nonhomogeneous Hidden Markov Model (NHMM) to decompose the sequences of merging processes as driving primitives. The model can consider the influence of exogenous factors on the evolution of merging behavior patterns and reveal the mechanism of pattern evolution.

3) Applying the time series K-means clustering to gather variable-length driving primitives into finite merging behavior patterns and explain them semantically.

The rest of this article is organized as follows. Section 2 documents the primitive based framework and its components. Section 3 introduces the datasets used in this study. Section 4 describes the experimental setup and the selection of optimal parameters. Section 5 presents the results and discussions. Section 6 provides the conclusions and future work for this study.

**METHODS**

The framework proposed in this study is shown in **Figure 1**, which consists of four steps: merging behavior extraction, primitive segmentation based on the NHMM, the Time Series K-means (TSKM) clustering, and evolutionary mechanism analysis. The following sections describe the NHMM and TSKM mathematically.

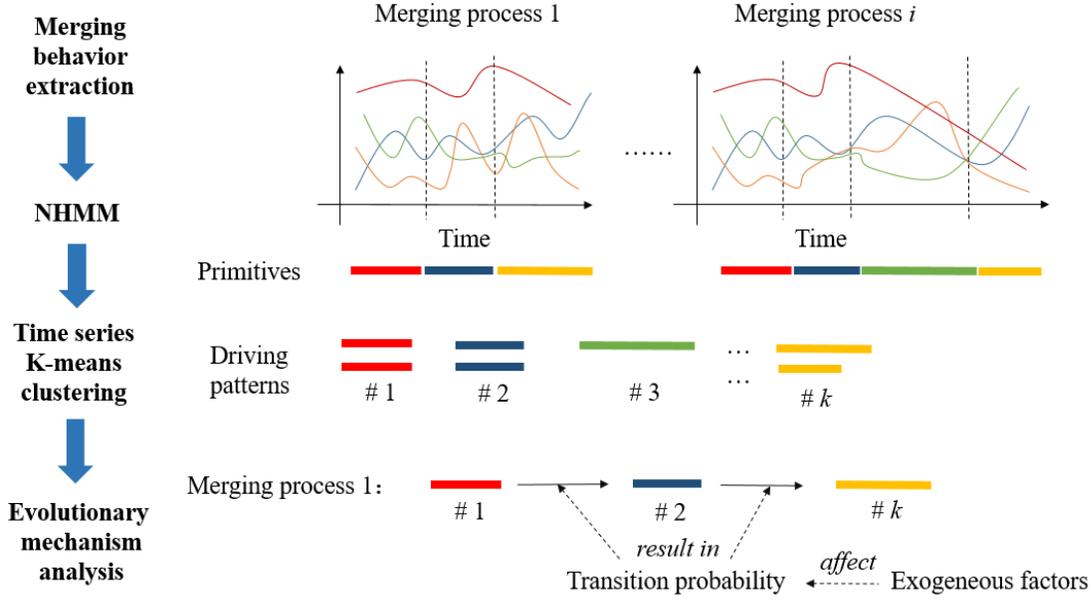

**Figure 1 Framework of merging behavior pattern recognition and evolutionary mechanism analysis**

**Merging behavior primitives**
The dynamic merging behavior could be reflected by movement parameters of the merging vehicles. The behavior shown at each moment constitutes the entire merging process O, expressed as Eq.(1).

$$O = \{o_1 \cdots, o_t \cdots, o_T\} \quad (1)$$

Where $o_t = \{v_x, v_y, acc_x, acc_y\} \in \mathbb{R}^6$ is the driver's merging behavior at time t. $v$ and $acc$ are velocity and acceleration of the merging vehicle, respectively. x and y indicate longitudinal and lateral, respectively. T represents the time length of a merging process. According to the HMM, the primitive of O could be decomposed into n segments called primitives (*20; 21*) as shownin Eq.(2).

$$O = \{p_1, \cdots, p_i, \cdots, p_n\} \quad (1 \leq n \leq T) \quad (2)$$

Where $p_i$ is the *i*-th primitive.

**Hidden Markov Model**
The HMM is a common approach to model the dynamic of multivariate time series $O = \{o_1 \cdots, o_t \cdots, o_T\}$, where the observations O are considered as a stochastic function of a hidden finite-state Markov process $q = \{q_1, q_2, \cdots, q_t\}$. Some other concepts and symbols involved in the HMM should be defined in advance.

Hidden states are expressed as $S = \{S_1, S_2, \cdots, S_N\}$, where N is the number of hidden states and $q_t \in S$ is the state at time t. Although the states are hidden, there are often physical meaning in applications *(18)*.

Emission probability is the conditional probability of an observation $o_t$ generated from state $q_t$, defined in Eq.(3).

$$b_t = P(o_t | q_t = k, \theta), k \in \{1, \cdots N\} \quad (3)$$

*Zhang et al.*

Transition probability represents the probability of transition from state *i* to state *j*. Assuming that $A = \{a_{ij}\}$ is the state transfer matrix. And the transfer probability for (*i,j*) pairs can be expressed as Eq.(4).

$$a_{ij} = P(q_{t+1} = S_j | q_t = S_i), \text{ s.t. } i \geq 1, j \leq N \text{ and } \sum_{j=1}^{N} a_{ij} = 1 \; \forall i \quad (4)$$

**Nonhomogeneous Hidden Markov Model**
The time homogeneity of the standard HMM limits its application in practice, for example, when $o_t$ is nonstationary. The NHMM relaxes this assumption and allow the transition probability of the hidden states to be dependent on exogenous variables. To introduce exogenous dependence into the transition matrix, the NHMM modulates the probability of entering each state using a set of K weighted regressors. The intuitive interpretation is the external variables control how likely the Markovian chain will enter state j via the logistic link and regression coefficients *(24)*. Thus, the transition probabilities change over time since the exogenous variable's value change over time. The time-varying transition probability matrix can reveal how the driving primitives change. The transition probability at time t are described using a multinomial logistic link function (Eq.(5))

$$a_{ijt} = P(q_t = S_j | q_{t-1} = S_i, x_t, \zeta) = \frac{exp(\xi_{ij} + x_t' \rho_j)}{\sum_{m=1}^{N} exp(\xi_{ij} + x_t' \rho_j)} \quad (5)$$

$x_t = \{\Delta x_f, \Delta x_r, \Delta x_{ft}, \Delta x_{rt}, l, d\}$ is a 6-dimensional exogenous covariate time series, where $\Delta x.$ is the longitudinal relative distance to the leading vehicle in the acceleration lane, $\{f, r, ft, rt\}$ represent surrounding vehicles around the merging vehicle, $l$ is the distance between the current location and the entrance of the ramp, and $d$ is density of the ramp (The data description section will explain these variables in more detail). $\rho_j$ is a vector of 6-dimensional coefficient of $x_t$, which indicates the influence of covariates on the probability of entering state j. For simplicity, we let $\zeta = \zeta_{ij} = (\rho_j, \xi_{ij})$ where $i, j \in \{1, \cdots N\}$. One of the $\zeta_{\cdot j}$ is set as 0 for identifiability. The entire process of the multivariate NHMM is illustrated in **Figure 2**. The gray boxes in **Figure 2** are observed values. The figure shows how the exogenous variable $x_t$ affects the model. Assuming that the state variable $q_t$ is known, the conditional likelihood value of the NHMM can be calculated by Eq.(6).

$$P(o_t | x, q, \zeta, \theta) = \prod_{t=1}^{T} f(o_t | q_t, \theta) P(q_t | q_{t-1}, x_t, \zeta) \quad (6)$$

Where $\theta$ is a set of parameters of the emission probability. When the latent state sequence is unknown, the likelihood for HMMs can be calculated by using the usual recursive method to marginalize over the unknown $q$ values.

*Zhang et al.*

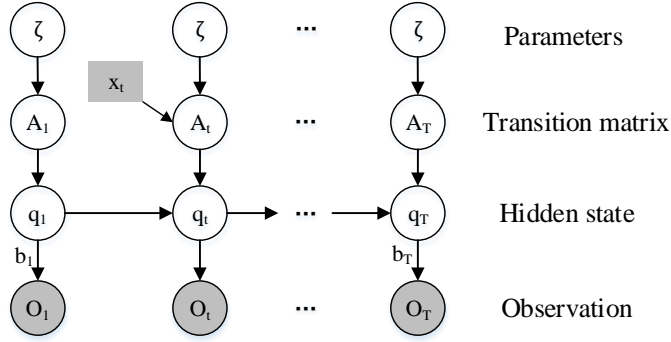

Figure 2 The graphical model of the NHMM

The decomposing of driving primitives mainly depends on hidden states of the Markov chain. In order to find the optimal hidden state sequence, this study adopts the Direct Gibbs (DG) sampler which requires more iterations but less cost per iteration. After identifying the optimal state sequence, the continuous observations with the same hidden state are segmented into driving primitives.

**Time series clustering**
The primitives from the NHMM are time series composed of the temporal features of merging behaviors. Abundant primitives may contain the same merging behavior patterns. In order to identify different merging behavior patterns, clustering method is used to gather similar primitives into a group. Note that these primitives are variable-length time series. The traditional clustering method converts the sequence length into the same by interpolation, but it will lose the duration attribute of primitives. For example, drivers maintain a state for 1 second and maintain the state for 10 seconds, which expresses different driving behavior patterns. Thus, the TSKM *(25)* is applied as the primitive clustering method. And the Dynamic Time Warping (DTW) is as the metric to compute the distance between primitives. DTW distance is the optimal alignment length of two primitives based on shape similarity. Assuming all primitives $P = \{p_1, p_2, \cdots p_n\}$ are gathered into K clusters $C = \{C_1, C_2, \cdots C_K\}$ and the center of each cluster is $\mu_i$. The objective is to minimize the within-cluster sum-of-squares in Eq.(7).

$$\lambda_w = min \sum_{i=1}^{K} \sum_{P \in C_i} \|p_i - \mu_i\|^2 \qquad (7)$$

**DATA DESCRIPTION**

**Real-World Dataset**
In order to extract driving patterns in the real world, this study selects the INTERACTION *(26)* dataset for merging process extraction and analysis. The INTERACTION dataset is an open data collected by Unmanned Aerial Vehicles (UAVs). Other common open datasets used to analyze micro driving behavior include: the Next Generation SIMulation (NGSIM) dataset *(27)*, highD dataset *(28)*, and Argoverse dataset *(29)*. Compared with these datasets, the INTERACTION dataset contains more merging scenarios with close interaction between vehicles at freeway on-ramps in congested traffic. This study selects the on-ramp merging scenario in China for analysis. As shown in Figure 3, the study area is bounded by the red line. The video length is 94.62 minutes and this data contain 10359 vehicles. The data collection frequency was 10 Hz.

*Zhang et al.*

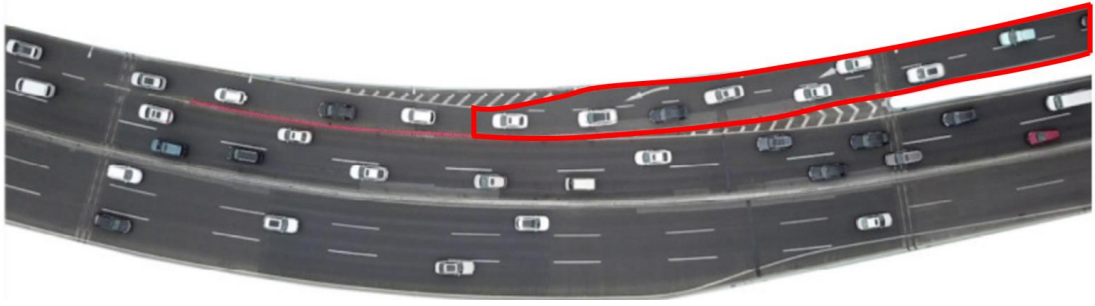

**Figure 3 The merging scenario at the freeway on-ramp in the INTERACTION dataset.**

**Data processing**

The merging process referred to in this study only considers the process of executing lane change, regardless of the gap selection process.

**Figure 4** displays the geometric characteristic and a merging process at the on-ramp.

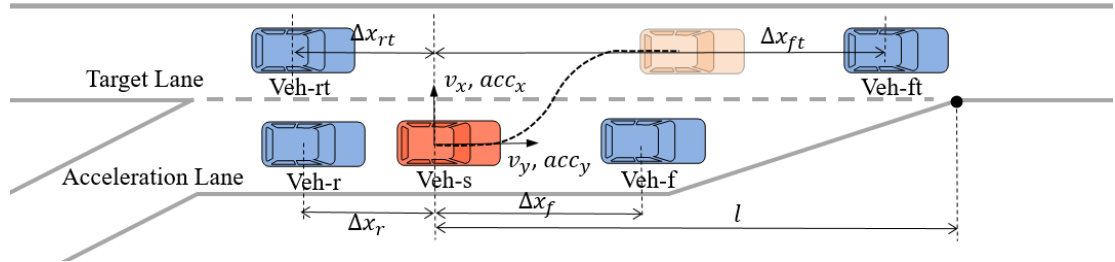

**Figure 4 Schematic diagram of a merging process at freeway on-ramp and the definitions of key variables**

The steps for extracting the merging processes and feature variables are as follows.

Step 1: Determine the time *t* when a vehicle crosses the lane boundary line.

Step 2: Find the starting and ending point of the lane changing processes forward and backward from time *t* respectively. Because the selected road section is not straight, the starting and ending point of the lane change cannot be simply judged by the thresholds of steering angle, lateral displacement or lateral velocity. After manually summarizing several merging processes, we obtain the criterion. That is, the two maximum points on the vehicles' lateral acceleration curve located before and after time t are the starting point and the ending point, respectively.

Step 3: Match the lead and lag vehicles closest to the merging vehicle in the acceleration lane and target lane, respectively. Calculate the longitudinal relative distance between the merging vehicle and these four vehicles at each moment. If no vehicle is detected at the location of some of the four vehicles, the relative distance is set to the length of target lane (This value is set as 130m).

Step 4: Calculate the values of feature variables characterizing driving behaviors and other covariates.

Step 5: Eliminate the merging process involving trucks. Since the proportion of trucks in the dataset is extremely low, it is not suitable to analyze the impact of vehicle types on the merging behavior. Therefore, this study focuses on the driving behavior patterns only involving cars during the merging processes.

In the end, a total of 813 merging processes are extracted for analysis in the following sections. A detailed description of all variables is shown in Table 1.

**TABLE 1 Descriptive statistics of variables used in this study**

| Variables | Description | Mean | S.D. | Min | Max |
|---|---|---|---|---|---|
| $acc_x$ | The lateral acceleration of the merging vehicle (m/s$^2$) | 0.06 | 0.45 | -3.52 | 2.07 |
| $acc_y$ | The longitudinal acceleration of the merging vehicle (m/s$^2$) | -0.02 | 0.11 | -1.17 | 0.61 |
| $v_x$ | The longitudinal speed of the merging vehicle (m/s) | 2.46 | 1.33 | 0.00 | 13.42 |
| $v_y$ | The lateral speed of the merging vehicle (m/s) | 0.33 | 0.30 | -1.83 | 2.68 |
| $\Delta x_f$ | The lead gap between merging vehicle and lead vehicle in the acceleration lane (m) | 53.04 | 57.27 | 0.01 | 130.00 |
| $\Delta x_r$ | The lag gap between lag vehicle and merging vehicle in the acceleration lane (m) | 29.35 | 41.07 | 0.00 | 130.00 |
| $\Delta x_{ft}$ | The lead gap between merging vehicle and lead vehicle in the target lane (m) | 6.35 | 7.33 | 0.00 | 130.00 |
| $\Delta x_{rt}$ | The lag gap between lag vehicle and merging vehicle in the target lane (m) | 15.01 | 32.09 | 0.00 | 130.00 |
| l | The remaining distance from the merging vehicle to the end of acceleration lane (m) | 37.15 | 20.52 | 0.00 | 76.14 |
| d | The density of the ramp (pcu/m/lane) | 0.04 | 0.01 | 0.01 | 0.09 |

**EXPERIMENT AND PARAMETERS SETTING**

**Driving primitives segmentation**
In the process of segmenting the primitives of each merging process, the optimal parameter value of K needs to be determined. As the value of K increases, the parameters to be estimated in the model will also increase, which in turn leads to an increase in the model's complexity. Therefore, this study choose the Bayesian Information Criterion (BIC) value, which adds a penalty term for the number of parameters as the criterion for model selection The maximum value of K is set to 6, because too many states make it difficult for the parameter estimation of the model to converge. A process of selecting the optimal K value is shown in **Table 2**, where K=3 is selected as the optimal state number.

Then the merging process could be segmented into 3 primitives as **Figure 5**, where each primitive contains different semantic information. Primitive 1 displays that at the beginning of the merging process, the merging vehicle generates a lane-changing motivation, and then reduce the lateral speed and longitudinal speed respectively to prepare for lane change. In this segment, the lateral acceleration changes slightly, while the longitudinal acceleration changes in a regular sinusoidal curve around -1.2m/s$^2$. Primitive 2 describes the process of the merging vehicle waiting for the time to change lanes. The lateral speed remains constant, and the longitudinal direction starts to accelerate slowly. Until primitive 3, the vehicle finds the right time and quickly completes the merging maneuver in the direction of the target lane.

**TABLE 2 Select the optimal parameter K value in NHMM**

| K | 1 | 2 | 3 | 4 | 5 | 6 |
|---|---|---|---|---|---|---|
| **BIC** | 759.16 | 698.03 | **695.46** | 775.24 | 864.18 | 973.58 |

*Zhang et al.*

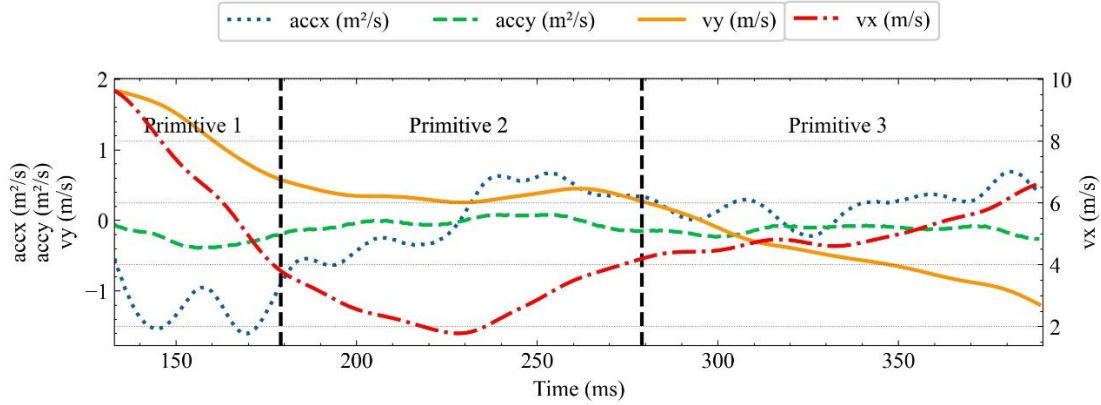

**Figure 5 Segmentation results of a merging process**

Implementing the NHMM to the 813 merging processes generates 4400 primitives. Inevitably, there will be some primitives with short periods which provide limited information. Thus, this study only retains primitives with the duration greater than 10 (1s at a sampling frequency of 10Hz).

**Merging behavior patterns clustering**
A large number of primitive segments contain the same merging behavior patterns. The TSKM algorithm gathers numerous primitives into finite merging behavior patterns existing in the real world. The number of clusters in the TSKM is determined based on the within-cluster sum-of-squares ($\lambda_w$) criterion. $\lambda_w$ is commonly used as the criteria to measure the clustering performance of the model, which measures the distance between samples in the same cluster. The smaller the value is, the better clustering performance the model has. In order to show the change of $\lambda_w$ more intuitively, the change rate curve of $\lambda_w$ is also shown in

**Figure 6**. According to the elbow rule, the optimal value of k is between 5 and 10. Combined with the change rate curve, this study choosee k=9 as the optimal number of clusters, where the change rate curve just starts to converge. Finally, all primitives of merging processes are grouped into 9 merging behavior patterns.

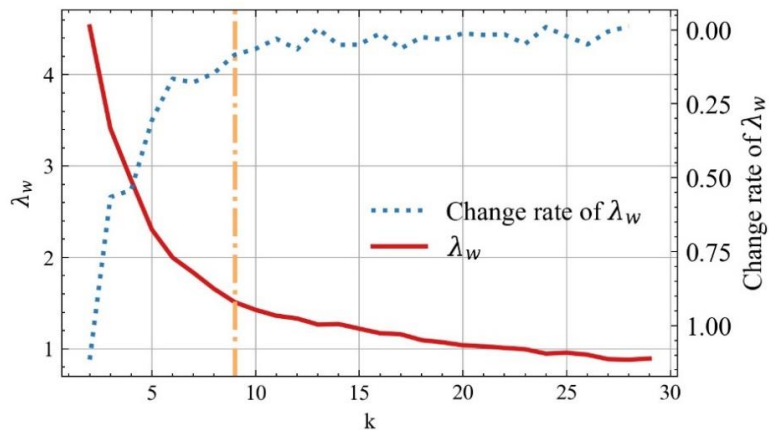

**Figure 6 The curve of $\lambda_w$ and its change rate over the number of clusters k**

**RESULTS AND DISCUSSION**

**Semantic representation of merging behavior patterns**

**Figure 7** shows the frequency distribution of all 9 patterns during the merging processes. In **Figure 7**, it can be seen that patterns #2 and #7 are the most common merging behavior patterns. Pattern #4 is the least common merging behavior patterns followed by patterns #6, #8, and #9.

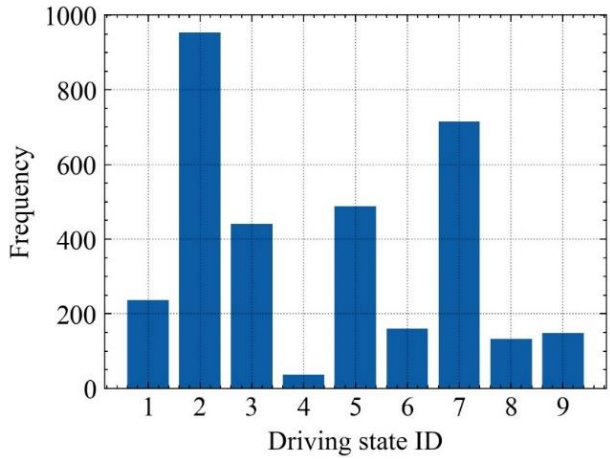

**Figure 7 The frequency distribution of each cluster of merging behavior patterns**

In general, cluster centroids are used to characterize the overall characteristics of each cluster. The cluster centroid of the merging behavior pattern is a 4-dimensional time series. In order to facilitate the semantic interpretation, this study uses the mean value of the cluster centroid to characterize each type of merging behavior. **Figure 8** intuitively displays all merging behavior patterns distinguished by colors, arrows and heading directions. The darker the red system, the greater the value in the positive direction, and the darker the blue system, the greater the value in the negative direction. Note that the comparison of color and arrow size only exists in the same variable, and there is no comparison between different variables. The arrow direction indicates the positive and negative direction of the value, and the arrow size indicates the size of the value. The heading direction is the combined speed of the longitudinal and lateral directions. Next, a detailed semantic explanation will be given to each merging behavior pattern.

Pattern #1 and pattern #3 represent the merging vehicle drives in the direction of the target lane and the acceleration direction promotes the merging process without being hindered by the surrounding environment. In Pattern #1, the vehicle has acceleration in the longitudinal direction, and basically maintains a constant speed in the lateral direction, which mainly occurs at the front section of the acceleration lane (a place far from the entrance). Pattern #3 shows that the vehicle has a low speed in both the longitudinal and lateral directions, but continues to accelerate in both directions.

Pattern #2 and pattern #7 display the vehicle travels at a very low speed, which usually occurs in an oversaturated state. The lateral speed of pattern #7 is very small without acceleration, indicating that the vehicle's willingness to merge is not strong. This merging behavior pattern usually occurs during the ramp queuing process and the merging vehicle is located at the back of the queue. Compared to pattern #7, pattern #2 has a larger velocity, especially in the lateral direction.

Pattern #4 describes a case, in which the merging vehicle moves into the target lane at a large velocity. Because the lane-changing conditions are not met, the vehicle decelerates rapidly in order to avoid collision. Similarly, pattern #9 also describes such a merging behavior pattern, but its average speed is half lower than pattern #4. They all show a strong willingness to merge, but the surrounding environment does not meet the requirements of merging. These two types of patterns are also called aggressive behaviors in some other studies.

In pattern #5, although the heading direction of the merging vehicle is facing to the target lane, it decelerates with small accelerations both longitudinally and laterally. This merging behavior pattern reflects the driver's conservative driving style.

Pattern #6 describes a special merging behavior pattern, that is, the heading direction is opposite to the merging path direction. The longitudinal speed of the vehicle continues to increase, while the lateral speed continues to increase in the negative direction. It usually occurs in the adjustment phase after entering the target lane. The lateral speed direction of pattern #8 is opposite to that of pattern #6, and the acceleration and deceleration behavior is more radical than that of pattern #6. Pattern #8 represents two cases. One is that the vehicle has made a merging action. Because the merging conditions are not met, the vehicle adjusts the lateral speed and continues to accelerate longitudinally to find the next lane-changing opportunity. The other is that the vehicle is in the adjustment stage just entering the target lane.

Examples and dynamic evolution of these behavior patterns in each specific merging process are shown in the following sections. These interpretable merging behavior patterns can used as labels of primitives to investigate the pattern transition during merging processes.

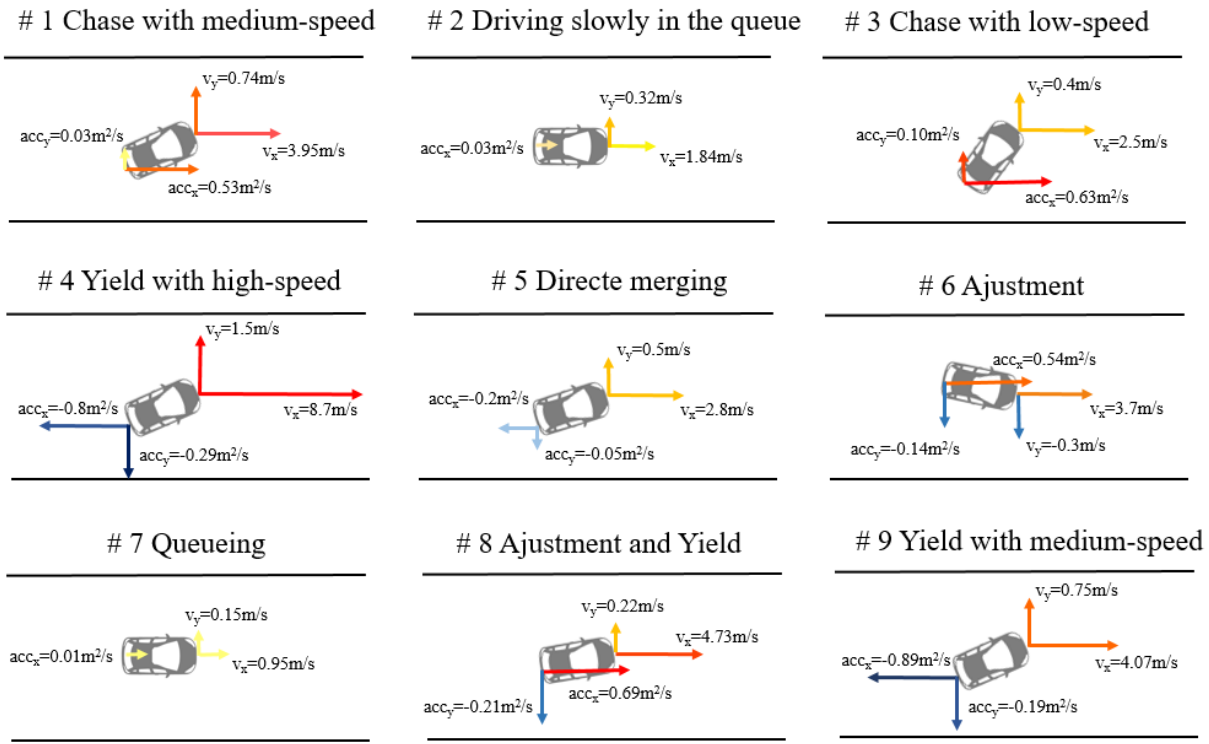

**Figure 8 Schematic diagram of 9 merging behavior patterns obtained after clustering**

**Overall influence of covariates**

As described in section NHMM, 6 exogenous covariates are included, which affect transition of merging behavior patterns. For all 813 merging processes, the statistically significant frequency of each covariate is shown in **Table 3**. Due to the heterogeneity of drivers, the transition probability of the merging behavior patterns depends on different covariates to varying degrees. Among them, the influence of $\Delta x_{ft}$ are highly significant for more than half of the merging events. However, less than 1/3 of the transition probability of merging events are affected by ramp density.

**TABLE 3 Statistically significant frequency of covariates**

| Covariates | $\Delta x_f$ | $\Delta x_r$ | $\Delta x_{ft}$ | $\Delta x_{rt}$ | $l$ | $d$ |
|---|---|---|---|---|---|---|
| Frequency | 354 | 341 | 433 | 303 | 358 | 264 |

**Transition of merging behavior patterns**

Each merging process is composed of these 9 behavior patterns, and different merging processes may contain different patterns. We select a merging process with similar patterns and a merging process with large differences in patterns for in-depth analysis.

    **Figure 9** and **Figure 10** show examples of merging behavior patterns evolution for two merging process. In each figure, sub-figure (a) displays the dynamic changes of feature variables characterizing driving behavior, the result of primitive segmentation and the evolution of the corresponding behavior pattern during a merging process. For example, in **Figure 9**, the whole merging process is segmented into three primitives. The values of sub-figure (b)-(c) and Table 4TABLE **5** are calculated based on primitives generated from the NHMM. According to the clustering results, we correspond these primitives to merging behavior patterns (**Figure 8**) for semantic interpretation and the patterns' ID are marked in parentheses in sub-figure (a). Sub-figure (b) shows the average value of the transition probability between each driving pattern of each moment. Sub-figure (c) describes in detail the transition probabilities value between every two patterns at each moment. Each subgraph in sub-figure (c) shows the time-varying transition probabilities at the corresponding position in sub-figure (b). **Table 4TABLE 5** shows the coefficients of the exogenous variables $x$ which control the transition of merging behavior patterns. There are K-1 coefficients for each exogenous variable because the k-th state in the NHMM is set to zero for indentifiability.

    **Figure 9** shows a merging process in a congested traffic flow, which includes the most common patterns #2 and #7. Primitives 1-3 corresponds to merging behavior patterns #7, #2 and #1, respectively. At the beginning of the merging process, the vehicle first drives slowly in pattern #2 and then turns to pattern #7 with a lower speed. In the oversaturated traffic condition, the vehicle has little space for acceleration and deceleration. Until merging conditions are met, the merging behavior switches to pattern #1 and quickly drives into the target lane. According to **Figure 9** (b), primitive 2 has the highest self-transition probability, and the probability of other primitives entering primitive 2 is relatively high, which indicates that the driver tends to maintain a stable and slow behavior pattern (pattern #2) for a long time. The transition between primitives at each moment is directly determined by the time-varying transition probability in **Figure 9** (c). First, the merging process starts from primitive 2. The self-transition probability of primitive 2 is close to 1 for a period of time, and primitive2 is maintained. During this period, the merging behavior pattern is pattern #2. With the change of time, the probability of primitive 2 to 1 increases sharply (t=200). The Markov chain of merging behavior switches to primitive1, and the driving pattern also changes from pattern #2 to pattern #7. Until t is close to 300ms, the probability from primitive1 to 3 gradually increases. The merging behavior finally transfer to primitive 3 (pattern #1). The factors controlling the transition of merging behavior patterns are given in **TABLE 4**. Each row in **TABLE 4** represents the influence of covariates on the probability of entering this primitive. For the first row, the smaller the gap to the veh-f and veh-ft, the more likely the merging behavior is to enter pattern #7. The small gap between vehicles implies that the traffic flow is in congested state. Thus, it is more likely to enter a slow driving pattern. The only factor that affects the probability of entering primitive 2 (pattern #2) is the distance between the veh-s and the veh-f ($\Delta x_f$). The smaller the value of $\Delta x_f$, the more likely the merging behavior is to enter pattern #2.

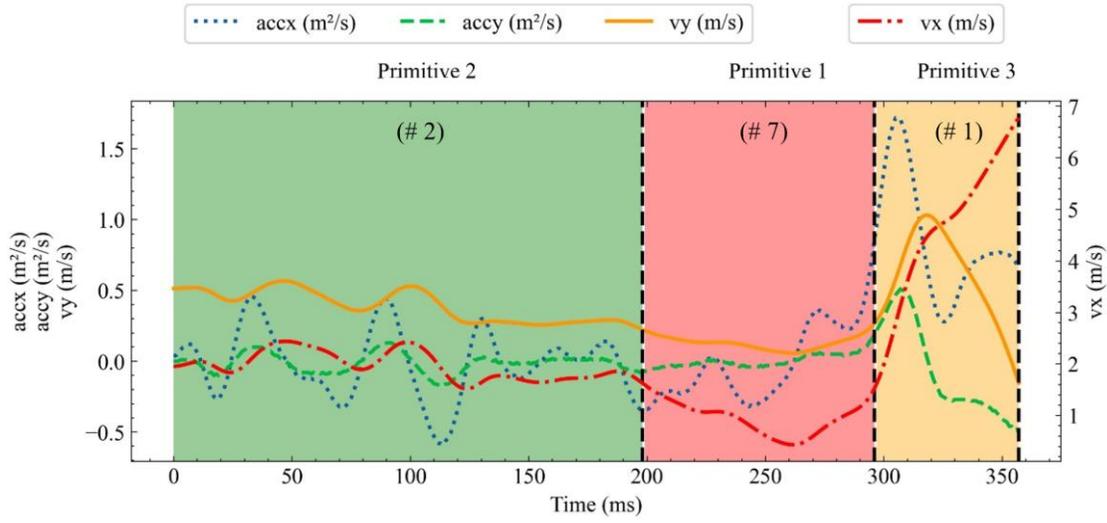

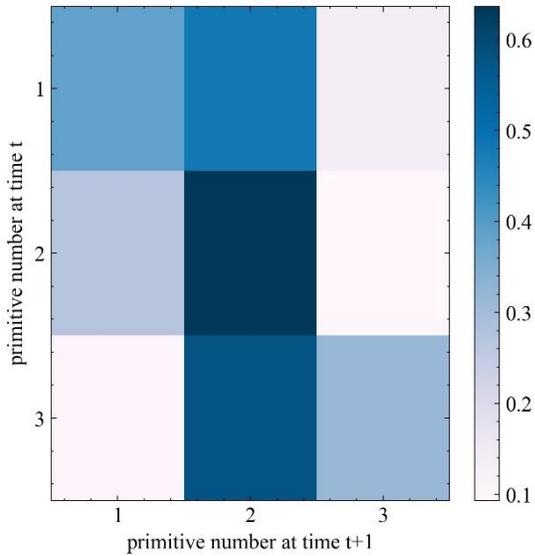
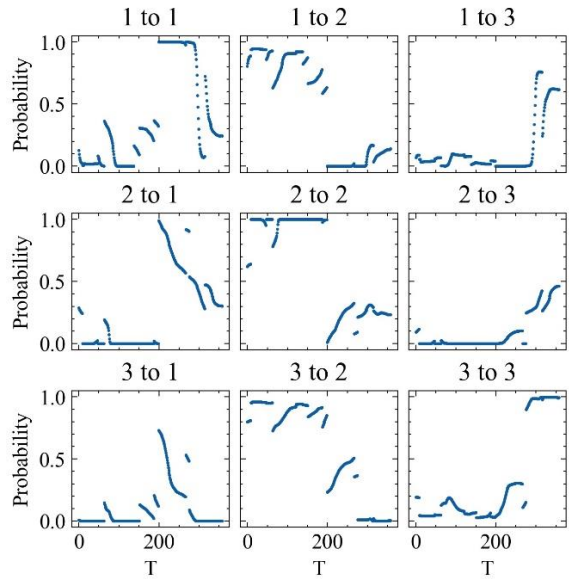

**Figure 9 The evolution of merging behavior patterns and the impact of exogenous factors for the merging process with similar patterns**

*Zhang et al.*

**TABLE 4 Coefficients of exogenous variables influencing the transition probabilities for the merging process with similar patterns**

| Covariates | $\Delta x_f$ | $\Delta x_r$ | $\Delta x_{ft}$ | $\Delta x_{rt}$ | $l$ | $d$ |
|---|---|---|---|---|---|---|
| Primitive 1 | **-71.85*** | 26.23 | **-107.10*** | 79.51 | -216.57 | -27.20 |
| Primitive 2 | **-130.13*** | -12.51 | 4.04 | -10.11 | 125.92 | -48.47 |

* statistically significant variables

**Figure 10** displays a merging process with three significantly different patterns. The merging process is segmented into three primitives associated with three patterns, noted as a pattern transition chain: #2→#3→#5. At the beginning, the vehicle drives slowly and steadily, then generates a strong willingness to merge and accelerates into the target lane. Finally, the vehicle slows down and adjusts slowly in both longitudinal and lateral directions. In **Figure 10**(b), in addition to the highest self-transition probability, the transition probability from primitive 1 to 3 is worth paying attention to (that is, pattern #2 to #3), which indicates that there is a high probability of merging with greater acceleration after a period of smooth and slow driving. The explanation of **Figure 10**(c) is similar to that of **Figure 9**(c). In **TABLE 5**, the first row shows the influence of six covariates on the probability of the merging behavior entering pattern #2, and the second row shows the effect on the entry to pattern #5. The driver's merging behavior is more likely to enter pattern #2 as the distance between veh-s and veh-rt ($\Delta x_{rt}$) becomes smaller. A small value of $\Delta x_{rt}$ means that the lane-changing conditions may not be satisfied, and the vehicle need to drive slowly and wait for the opportunity to change lanes. In the second row of **TABLE 5**, the smaller the distance between veh-s and the lead vehicles in the two lanes ($\Delta x_f$ and $\Delta x_{ft}$), the more likely it is to enter the pattern #5. The reason is that the merging vehicle needs to slow down and keep a safe distance from the lead vehicles. In addition, the probability that the merging behavior enters pattern #5 will increase as the distance from the merging entrance becomes smaller, because the vehicle needs to decelerate to ensure that it enters the target lane before reaching the end of acceleration lane.

*Zhang et al.*

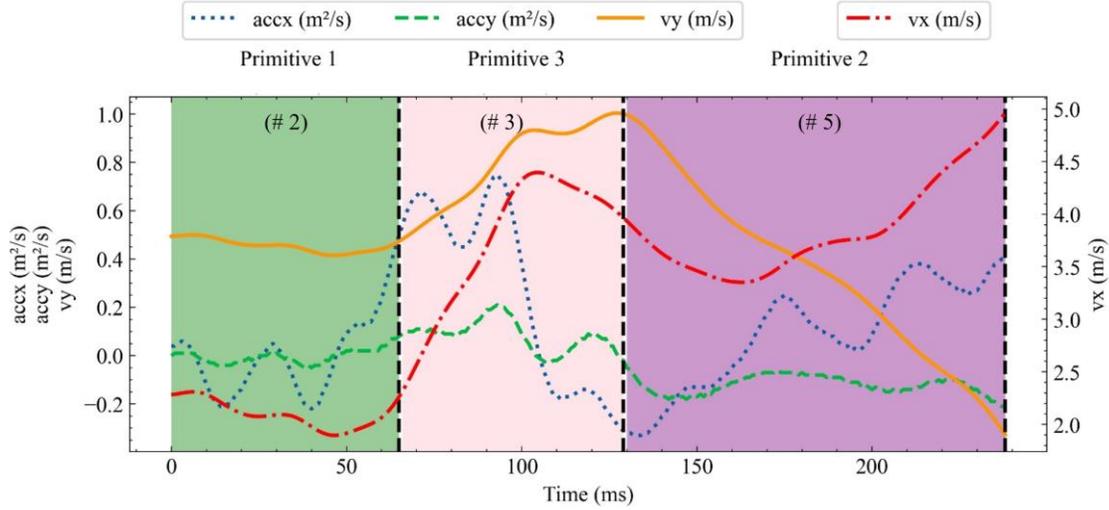

(a) Evolution of merging behavior patterns

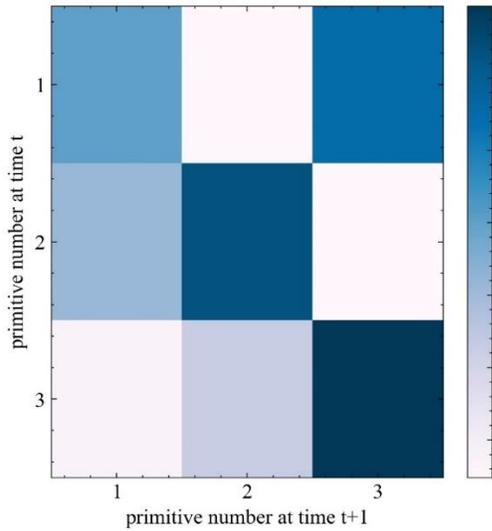

(b) tranaition matrix (mean)

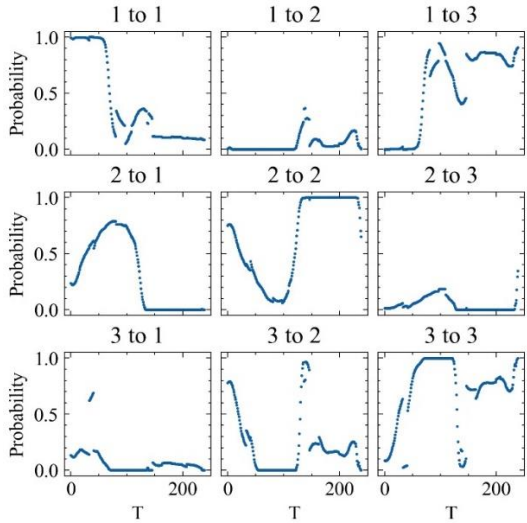

(c) tranaition matrix (values at each time)

**Figure 10 The evolution of merging behavior patterns and the impact of exogenous factors for the merging process with large differences in patterns**

**TABLE 5 Coefficients of exogenous variables influencing the transition probabilities for the merging process with large differences in patterns**

| Covariates | $\Delta x_f$ | $\Delta x_r$ | $\Delta x_{ft}$ | $\Delta x_{rt}$ | $l$ | $d$ |
|---|---|---|---|---|---|---|
| Primitive 1 | -122.29 | 62.39 | -50.66 | **-64.69*** | 11.75 | -6.43 |
| Primitive 2 | **-384.26*** | 30.96 | **-388.56*** | -56.18 | **-543.35*** | 9.32 |

## CONCLUSIONS

Understanding driving patterns based on real-world data can provide more prior knowledge and facilitate autonomous vehicles to make reasonable decisions. This study proposed a primitive based framework leveraging the continuous and discrete processes to learn underlying merging behavior patterns and investigate their evolution mechanism at freeway on-ramps in congested traffic flow. In the proposed framework, the Nonhomogeneous Hidden Markov Model (NHMM) was employed to decompose merging processes into several primitives considering the influences of exogenous factors. The time-series K-means clustering was developed to gather numerous primitives with variable-length time series into 9 semantically interpretable groups. A merging process consists of several combinations of these 9 types of patterns. The most common merging behavior patterns are patterns #2 and #7, indicating that most drivers in this place tend to drive slowly and smoothly. Furthermore, the proposed method also reveals the evolution mechanism of patterns and the influence of exogenous variables on pattern transition.

In the research process, it was found that there is significant heterogeneity in the merging processes. Different drivers' merging processes may be composed of different driving patterns. The proposed method in this study can also analyze the heterogeneity of drivers in the merging scenarios according to the evolution law of driving patterns. Since the data used in this paper do not include multiple merging processes of the same driver, this will become a future research direction. In addition, more exogenous factors affecting driving behaviors can be taken into account, such as weather, different ramp types, traffic incidents and driving environment in different traffic facilities.

## ACKNOWLEDGMENTS

This research was funded by the National Natural Science Foundation of China (Grant No. 71971160), the Shanghai Science and Technology Committee (Grant No. 19210745700) and the Fundamental Research Funds for the Central Universities (Grant No. 2120210009).

## AUTHOR CONTRIBUTIONS

The authors confirm contribution to the paper as follows: study conception and design: Yue Zhang and Yajie Zou; data collection: Yue Zhang and Wanbing Han; analysis and interpretation of results: Yue Zhang and Yajie Zou; draft manuscript preparation: Yue Zhang and Yajie Zou. All authors reviewed the results and approved the final version of the manuscript.

# REFERENCES

1. Zhao, L., J. Sun, and H. M. Zhang. Observations and analysis of multistep-approach lane changes at expressway merge bottlenecks in Shanghai, China. *Transportation Research Record: Journal of the Transportation Research Board*, 2013. 2395(1): 73-82.

2. Zhang, C., J. Zhu, W. Wang, and J. Xi. Spatiotemporal learning of multivehicle interaction patterns in lane-change scenarios. *IEEE Transactions on Intelligent Transportation Systems*, 2021.

3. Lee, G. Modeling gap acceptance at freeway merges. In, Massachusetts Institute of Technology, 2006.

4. Marczak, F., W. Daamen, and C. Buisson. Merging behaviour: Empirical comparison between two sites and new theory development. *Transportation Research Part C: Emerging Technologies*, 2013. 36: 530-546.

5. Toledo, T., H. N. Koutsopoulos, and M. Ben-Akiva. Estimation of an integrated driving behavior model. *Transportation Research Part C: Emerging Technologies*, 2009. 17(4): 365-380.

6. Gu, X., M. Abdel-Aty, Q. Xiang, Q. Cai, and J. Yuan. Utilizing UAV video data for in-depth analysis of drivers' crash risk at interchange merging areas. *Accident Analysis & Prevention*, 2019. 123: 159-169.

7. Sun, J., J. Ouyang, and J. Yang. Modeling and analysis of merging behavior at expressway on-ramp bottlenecks. *Transportation Research Record: Journal of the Transportation Research Board*, 2014. 2421(1): 74-81.

8. Chu, T. D., T. Miwa, and T. Morikawa. Discrete choice models for gap acceptance at urban expressway merge sections considering safety, road geometry, and traffic conditions. *Journal of Transportation Engineering, Part A: Systems*, 2017. 143(7): 04017025.

9. Wan, X., P. J. Jin, H. Gu, X. Chen, and B. Ran. Modeling freeway merging in a weaving section as a sequential decision-making process. *Journal of Transportation Engineering, Part A: Systems*, 2017. 143(5): 05017002.

10. Meng, Q., and J. Weng. An improved cellular automata model for heterogeneous work zone traffic. *Transportation Research Part C: Emerging Technologies*, 2011. 19(6): 1263-1275.

11. Weng, J., G. Du, D. Li, and Y. Yu. Time-varying mixed logit model for vehicle merging behavior in work zone merging areas. *Accident Analysis & Prevention*, 2018. 117: 328-339.

12. Arbis, D., and V. V. Dixit. Game theoretic model for lane changing: Incorporating conflict risks. *Accident Analysis & Prevention*, 2019. 125: 158-164.

13. Jing, S., F. Hui, X. Zhao, J. Rios-Torres, and A. J. Khattak. Cooperative Game Approach to Optimal Merging Sequence and on-Ramp Merging Control of Connected and Automated Vehicles. *IEEE Transactions on Intelligent Transportation Systems*, 2019. 20(11): 4234-4244.

14. Meng, Q., and J. Weng. Classification and regression tree approach for predicting drivers' merging behavior in short-term work zone merging areas. *Journal of Transportation Engineering*, 2012. 138(8): 1062-1070.

15. El Abidine, K. Z., A. Samir, and B. Rebiha. A Dynamic Bayesian Network Based Merge Mechanism for Autonomous Vehicles. In *Proceedings of the AAAI Conference on Artificial Intelligence, No. 33*, 2019.

9953-9954.

16.Tang, J., F. Liu, W. Zhang, R. Ke, and Y. Zou. Lane-changes prediction based on adaptive fuzzy neural network. *Expert Systems with Applications*, 2018. 91: 452-463.

17.Higgs, B., and M. Abbas. Segmentation and Clustering of Car-Following Behavior: Recognition of Driving Patterns. *IEEE Transactions on Intelligent Transportation Systems*, 2015. 16(1): 81-90.

18.Rabiner, L. R. A tutorial on hidden Markov models and selected applications in speech recognition. *Proceedings of the IEEE*, 1989. 77(2): 257-286.

19.Wang, W., J. Xi, and D. Zhao. Driving style analysis using primitive driving patterns with Bayesian nonparametric approaches. *IEEE Transactions on Intelligent Transportation Systems*, 2018. 20(8): 2986-2998.

20.Zhang, W., and W. Wang. Learning v2v interactive driving patterns at signalized intersections. *Transportation Research Part C: Emerging Technologies*, 2019. 108: 151-166.

21.Zhang, Y., Y. Zou, and L. Wu. V2V Spatiotemporal Interactive Pattern Recognition and Risk Analysis in Lane Changes. *arXiv preprint arXiv:2105.10688*, 2021.

22.Li, G. Application of finite mixture of logistic regression for heterogeneous merging behavior analysis. *Journal of Advanced Transportation*, 2018. 2018.

23.Li, G., and J. Cheng. Exploring the Effects of Traffic Density on Merging Behavior. *IEEE Access*, 2019. 7: 51608-51619.

24.Holsclaw, T., A. M. Greene, A. W. Robertson, and P. Smyth. Bayesian nonhomogeneous Markov models via Pólya-Gamma data augmentation with applications to rainfall modeling. *The Annals of Applied Statistics*, 2017. 11(1): 393-426.

25.Tavenard, R., J. Faouzi, G. Vandewiele, F. Divo, G. Androz, C. Holtz, M. Payne, R. Yurchak, M. Rußwurm, and K. Kolar. Tslearn, A Machine Learning Toolkit for Time Series Data. *Journal of Machine Learning Research*, 2020. 21(118): 1-6.

26.Zhan, W., L. Sun, D. Wang, H. Shi, A. Clausse, M. Naumann, J. Kummerle, H. Konigshof, C. Stiller, and A. de La Fortelle. Interaction dataset: An international, adversarial and cooperative motion dataset in interactive driving scenarios with semantic maps. *arXiv preprint arXiv:1910.03088*, 2019.

27.Alexiadis, V., J. Colyar, J. Halkias, R. Hranac, and G. McHale. The next generation simulation program. *Institute of Transportation Engineers. ITE Journal*, 2004. 74(8): 22.

28.Krajewski, R., J. Bock, L. Kloeker, and L. Eckstein. The highd dataset: A drone dataset of naturalistic vehicle trajectories on german highways for validation of highly automated driving systems.In *2018 21st International Conference on Intelligent Transportation Systems (ITSC)*, IEEE, 2018. 2118-2125.

29.Chang, M.-F., J. Lambert, P. Sangkloy, J. Singh, S. Bak, A. Hartnett, D. Wang, P. Carr, S. Lucey, and D. Ramanan. Argoverse: 3d tracking and forecasting with rich maps.In *Proceedings of the IEEE/CVF Conference on Computer Vision and Pattern Recognition*, 2019. 8748-8757.